\documentstyle[psfig,epsfig,aps]{revtex}
\def\eq#1{{eq.~(\ref{#1})}}

\def\beq{\begin{equation}}
\def\eeq{\end{equation}}
\def\lsim{\raise0.3ex\hbox{$\;<$\kern-0.75em\raise-1.1ex\hbox{$\sim\;$}}}
\def\gsim{\raise0.3ex\hbox{$\;>$\kern-0.75em\raise-1.1ex\hbox{$\sim\;$}}}

\def\aa#1#2#3{ Astron. \& Astroph. {\bf #1}, #3 (19#2)}
\def\aj#1#2#3{ Astron. J. {\bf #1}, #3 (19#2)}
\def\apj#1#2#3{ Astrophys. J. {\bf #1}, #3 (19#2)}

\def\mn#1#2#3{ MNRAS {\bf #1}, #3 (19#2)}
 
\def\np#1#2#3{ Nucl. Phys. {\bf #1}, #3 (19#2)}
\def\nat#1#2#3{ Nature {\bf #1}, #3 (19#2)}
\def\pl#1#2#3{ Phys. Lett. {\bf #1}, #3 (19#2)}
\def\pr#1#2#3{ Phys. Rev. {\bf #1}, #3 (19#2)}

\def\prl#1#2#3{ Phys. Rev. Lett. {\bf #1}, #3 (19#2)}

\def\rpp#1#2#3{ Rept. Prog. Phys. {\bf #1}, #3 (19#2)}

\newcommand{\ch}{\mathrm{ch}}

\def\Frac#1#2{\frac{\displaystyle{#1}}{\displaystyle{#2}}}
\begin{document}
\title{Cosmological implications of a Relic Neutrino Asymmetry}
\author{Julien Lesgourgues~\footnote{ E-mail: lesgour@sissa.it}
and Sergio Pastor~\footnote{ E-mail: pastor@sissa.it}}
\address{SISSA--ISAS and INFN, Sezione di Trieste\\
Via Beirut 2-4,I-34013 Trieste, ITALY}
\maketitle
\begin{abstract}
We consider some consequences of the presence of a cosmological lepton
asymmetry in the form of neutrinos. A relic neutrino degeneracy enhances
the contribution of massive neutrinos to the present energy density of
the Universe, and modifies the power spectrum of radiation and matter.
Comparing with current observations of cosmic microwave background
anisotropies and large scale structure, we derive some constraints on
the neutrino degeneracy and on the spectral index in the case of a
flat Universe with a cosmological constant.
\end{abstract}

\section{Introduction}

It is generally assumed that our Universe contains an approximately
equal amount of leptons and antileptons. The lepton asymmetry would be
of the same order as the baryon asymmetry, which is very small as
required by Big Bang Nucleosynthesis (BBN) considerations.  The
existence of a large lepton asymmetry is restricted to be in the form
of neutrinos from the requirement of universal electric neutrality,
and the possibility of a large neutrino asymmetry is still open. {}From
a particle physics point of view, a lepton asymmetry can be generated
by an Affleck-Dine mechanism \cite{AF} without producing a large
baryon asymmetry (see ref.~\cite{Casas} for a recent model), or even
by active-sterile neutrino oscillations after the electroweak phase
transition (but in this last case, it might not be of order unity)
\cite{Foot}. This lepton asymmetry can postpone symmetry restoration
in non-supersymmetric or supersymmetric models \cite{Linde} (note
that this is also true for other charges \cite{Riotto}).

In this paper we study some cosmological implications of relic
degenerate neutrinos\footnote{Here, by degeneracy, we mean that there exists a
large neutrino-antineutrino asymmetry, or vice versa, and not a
degeneracy in the mass sense.}. We do not consider any specific model
for generating such an asymmetry, and just assume that it was created
well before neutrinos decouple from the rest of the plasma.  An
asymmetry of order one or larger can have crucial effects on the
global evolution of the Universe. Among other effects, it changes the
decoupling temperature of neutrinos, the primordial production of
light elements at BBN, the time of equality between radiation and
matter, or the contribution of relic neutrinos to the present energy
density of the Universe. The latter changes affect the evolution of
perturbations in the Universe. We focus on the anisotropies of the
Cosmic Microwave Background (CMB), and on the distribution of Large
Scale Structure (LSS). We calculate the power spectrum of both
quantities, in the case of massless degenerate neutrinos, and also for
neutrinos with a mass of $0.07$ eV, as suggested to explain the
experimental evidence of atmospheric neutrino oscillations at
Super-Kamiokande \cite{SK}.  The cosmological implications of
neutrinos with such a small mass are known to be very small,
but we will see that this conclusion is modified if a large 
neutrino degeneracy exists.  We also include in our analysis the
possibility that the dominant contribution to the present energy
density in the Universe is due to a cosmological constant:
$\Omega_\Lambda \sim 0.6-0.7$, keeping the Universe flat
($\Omega_0+\Omega_\Lambda=1$), as suggested by recent observations
(see \cite{Lineweaver} and references therein).

The effect of neutrino degeneracy on the LSS power spectrum was
studied in ref.~\cite{Larsen}, as a way of improving the agreement
with observations of mixed dark matter models with eV neutrinos, in
the case of high values of the Hubble parameter.  Also, Adams \&
Sarkar \cite{Sarkar} calculated the CMB anisotropies and the matter
power spectrum, and compared them with observations in the
$\Omega_\Lambda=0$ case for massless degenerate neutrinos. More
recently, Kinney \& Riotto \cite{Kinney} also calculated the CMB
anisotropies for massless degenerate neutrinos in the
$\Omega_\Lambda=0.7$ case.

This paper is organized as follows. In section \ref{energy}, we
calculate the contribution of massive degenerate neutrinos to the
present energy density of the Universe.  In section \ref{power}, we
explain how to calculate the power spectra, with the help of the code
{\tt cmbfast} \cite{SelZal}. In section \ref{results}, we discuss the
effect of the degeneracy on CMB anisotropies and the matter power
spectrum, both for massless neutrinos and $m_{\nu}=0.07$ eV.  Finally,
in section \ref{comparison}, we derive some constraints on the
neutrino degeneracy from CMB and LSS data, in the particular case of a
flat Universe with an arbitrary cosmological constant and for standard
values of other cosmological parameters.

\section{Energy density of massive degenerate neutrinos}
\label{energy}

The energy density of one species of massive degenerate neutrinos and
antineutrinos, described by the distribution functions $f_\nu$ and
$f_{\bar{\nu}}$, is (here and throughout the paper we use
$\hbar=c=k_B=1$ units)
\beq
\rho_\nu + \rho_{\bar{\nu}}= 
\int \frac{d^3\vec{p}}{(2\pi)^3} ~E_\nu (f_\nu(p)+ f_{\bar{\nu}}(p))=
\frac{1}{2\pi^2}\int_0^\infty dp ~p^2 \sqrt{p^2 + m_\nu^2}
(f_\nu(p)+ f_{\bar{\nu}}(p))~,
\label{defrhonu}
\eeq
valid at any moment. Here $p$ is the magnitude of the 3-momentum and
$m_\nu$ is the neutrino mass.

When the early Universe was hot enough, the neutrinos were in
equilibrium with the rest of the plasma via the weak interactions. In
that case the distribution functions $f_\nu$ and $f_{\bar{\nu}}$
changed with the Universe expansion, keeping the form of a Fermi-Dirac
distribution,
\beq 
f_\nu(p)=\Frac{1}{\exp \left(\frac{p}{T_\nu} -
\frac{\mu}{T_\nu}\right)+1} \qquad 
f_{\bar{\nu}}(p)=\Frac{1}{\exp
\left(\frac{p}{T_\nu} + \frac{\mu}{T_\nu}\right)+1}
\label{FD}
\eeq
Here $\mu$ is the neutrino chemical potential, which is nonzero if a
neutrino-antineutrino asymmetry has been previously produced. Later
the neutrinos decoupled when they were still
relativistic\footnote{Unless the neutrino mass is comparable to the
decoupling temperature, ${\cal O}(m_\nu) \sim$ MeV.}, and from that
moment the neutrino momenta just changed according to the cosmological
redshift.  If $a$ is the expansion factor of the Universe, the
neutrino momentum decreases keeping $ap$ constant.  At the same time
the neutrino degeneracy parameter $\xi \equiv \mu/T_\nu$ is conserved,
with a value equal to that at the moment of decoupling. Therefore one
can still calculate the energy density of neutrinos now from
\eq{defrhonu} and \eq{FD}, replacing $\mu/T_\nu$ by $\xi$ and
$p/T_\nu$ by $p/(y_\nu T_0)$, where $T_0 \simeq 2.726$ K and $y_\nu$
is the present ratio of neutrino and photon temperatures, which is not
unity because once decoupled the neutrinos did not share the entropy
transfer to photons from the successive particle annihilations that
occurred in the early Universe. In the standard case, the massless
non-degenerate neutrinos decoupled just before the electron-positron
pairs annihilated to photons, from which one gets the standard factor
$y_\nu=(4/11)^{1/3}$.

In the presence of a significant neutrino degeneracy $\xi$ the
decoupling temperature $T(\xi)$ is higher than in the standard
case \cite{Freese,Kang}. The reaction rate $\Gamma$ of the weak
processes, that keep the neutrinos in equilibrium with the other
species, is reduced because some of the initial or final neutrino
states will be occupied. The authors of ref.~\cite{Kang} used the
Boltzmann equation to calculate $\Gamma$ for the process $\nu_d +
\bar{\nu}_d \leftrightarrow e^+ + e^-$ (here $\nu_d$ denotes
degenerate neutrinos), including the corresponding Fermi blocking
factors. It was found that the neutrino decoupling temperature is
$T_{dec}(\xi) \approx 0.2\xi^{2/3}\exp(\xi/3)$ MeV (for
$\nu_\mu$ or $\nu_\tau$). Therefore if $\xi$ is large enough,
the degenerate neutrinos decouple before the temperature of the
Universe drops below the different mass thresholds, and are not
heated by the particle-antiparticle annihilations.  The ratio of
neutrino and photon temperatures is thus reduced accordingly.

The present contribution of these degenerate neutrinos to the energy
density of the Universe can be parametrized as $\rho_\nu = 10^4 h^2
\Omega_\nu$ eV cm$^{-3}$, where $\Omega_\nu$ is the neutrino energy
density in units of the critical density $\rho_c=3H^2M_P^2/8\pi$,
$M_P=1.22 \times 10^{19}$ GeV is the Planck mass and $H=100h$ Km
s$^{-1}$ Mpc$^{-1}$ is the Hubble parameter.
The value of $\rho_\nu$ can be calculated as a function of the
neutrino mass and the neutrino degeneracy $\xi$, or equivalently the
present neutrino asymmetry $L_\nu$ defined as the following ratio of
number densities
\beq 
L_\nu \equiv \frac{n_\nu-n_{\bar{\nu}}}{n_\gamma} =
\frac{1}{12\zeta (3)} y^3_\nu [\xi^3 + \pi^2 \xi]
\label{Lnu}
\eeq
We show\footnote{Here we assume $\xi>0$, but the results are also
valid for $\xi<0$ provided that $\xi$ and $L_\nu$ are understood as
moduli.} in figures \ref{ximass} and \ref{lnumass} the contours in the
$(m_\nu,L_\nu)$ and $(m_\nu,\xi)$ planes that correspond to some
particular values of $h^2 \Omega_\nu$.  One can see from the figures
that there are two limits: massive non-degenerate neutrinos and
massless degenerate neutrinos. The first case corresponds to the
vertical lines, when one recovers the well-known bound on the neutrino
mass $m_\nu \lsim 46$ eV for $h^2 \Omega_\nu=0.5$. On the other hand,
for very light neutrinos, the horizontal lines set a maximum value on the
neutrino degeneracy, that would correspond to a present neutrino
chemical potential $\mu_0 \lsim 7.4 \times 10^{-3}$ eV, also for $h^2
\Omega_\nu=0.5$. In the intermediate region of the figures the
neutrino energy density is $\rho_\nu \simeq m_\nu n_\nu (\xi)$ and the
contours follow roughly the relations
$$
L_\nu \left (\frac{m_\nu}{\mbox{eV}} \right )\simeq 24.2 h^2\Omega_\nu
$$
\beq
(\pi^2\xi + \xi^3) \left (\frac{m_\nu}{\mbox{eV}} \right )
\simeq \frac{350}{y_\nu^3} h^2\Omega_\nu
\label{lines2}
\eeq

A similar calculation has been recently performed in reference
\cite{PalKar}. However the difference between neutrino and photon
temperatures was not properly taken into account for large $\xi$. 
It was argued that, since the number density of highly degenerate neutrinos is
larger than in the non-degenerate case, the neutrinos would have been longer in 
thermal contact with $e^+e^-$, therefore sharing with photons the entropy release.
However this is not the case \cite{Kang} as we discussed before.

The presence of a neutrino degeneracy can modify the outcome of Big
Bang Nucleosynthesis (for a review see \cite{Sarkar96}). First a
larger neutrino energy density increases the expansion rate of the
Universe, thus enhancing the primordial abundance of $^4$He. This is
valid for a nonzero $\xi$ of any neutrino flavor.  In addition if the
degenerate neutrinos are of electron type, they have a direct
influence over the weak processes that interconvert neutrons and
protons. This last effect depends on the sign of $\xi_{\nu_e}$. Both
effects may be simultaneously important and it could be possible in
principle to explain the observed primordial abundances with a large
baryon density, $\Omega_B h^2 \approx 1$ \cite{Freese,Kang}.  However
this possibility is ruled out by the fact that in that case our
Universe would have been radiation dominated during a longer period
and the observed large-scale structure would be difficult to
explain. {}From BBN one gets the following constraint \cite{Kang}
\beq
-0.06 \lsim \xi_{\nu_e} \lsim 1.1
\label{bbn}
\eeq
while a sufficiently long matter dominated epoch requires  
\beq
|\xi_{\nu_\mu,\nu_\tau}| \lsim 6.9
\label{lsf}
\eeq
This estimate from \cite{Kang} agrees with our analysis in section
\ref{comparison}. Assuming that the degenerate neutrinos are
$\nu_\mu$ or $\nu_\tau$, this places a limit on the degeneracy as shown by
the horizontal line in figures \ref{ximass} and \ref{lnumass}.
%
%

\section{Power spectra calculation}
\label{power}

We compute the power spectra of CMB anisotropies and large-scale
structure using the Boltzmann code {\tt cmbfast} by Seljak \&
Zaldarriaga \cite{SelZal}, adapted to the case of one family of
degenerate neutrinos ($\nu$, $\bar{\nu}$), with mass $m_\nu$ and
degeneracy parameter $\xi$.  Let us first review the required
modifications.  We use the notations of Ma \& Bertschinger
\cite{MaBer}, and for all issues not specific to our case, we refer
the reader to this review.

Background quantities can be rewritten in terms of two dimensionless
parameters ($M$, $Q$)
\begin{equation} \label{M}
M = \frac{m_\nu}{T_{\nu 0}}=
\frac{m_\nu(\mbox{eV})}{8.6170 \times 10^{-5}\times(4/11)^{1/3}~T_0(\mbox{K})},
\qquad 
Q = \frac{a~p}{T_{\nu 0}}
\end{equation}
(we are assuming $y_\nu = (4/11)^{1/3}$, and therefore $\xi \leq 12$ \cite{Kang};
the scale factor is defined so that $a = 1$ today).  For
Super-Kamiokande neutrinos with $m_\nu=0.07$ eV, $M \simeq 417$.  We
then get for the mean density, pressure and phase-space distributions
\begin{eqnarray} \label{background}
\bar{\rho}_\nu + \bar{\rho}_{\bar{\nu}}
&=& 
\frac{T^4_{\nu}}{2 \pi^2} 
\int Q^2 dQ \sqrt{Q^2 + a^2 M^2}
\left( f_{\nu} (Q) + f_{\bar{\nu}} (Q) \right), 
\nonumber \\ 
\bar{P}_\nu + \bar{P}_{\bar{\nu}}
&=& 
\frac{T^4_{\nu}}{6 \pi^2} 
\int Q^2 dQ 
\frac{Q^2}{\sqrt{Q^2 + a^2 M^2}}
\left( f_{\nu} (Q) + f_{\bar{\nu}} (Q) \right), 
\\
f_{\nu}(Q)  &=& \frac{1}{e^{Q-\xi}+1}, \qquad
f_{\bar{\nu}}(Q) = \frac{1}{e^{Q+\xi}+1}. \nonumber
\end{eqnarray}
In the case of massive degenerate neutrinos, these integrals must be
calculated for each value of the scale factor, and also at the
beginning of the code in order to find $\Omega_{\nu}$ today. On the
other hand, for massless neutrinos, there is an exact analytic
solution
\begin{equation}
\bar{\rho}_{\nu} + \bar{\rho}_{\bar{\nu}}
= 3 (\bar{P}_{\nu} + \bar{P}_{\bar{\nu}} )
= \frac{7}{8} 
\frac{\pi^2}{15} 
T^4_{\nu}
\left[ 1+
\frac{30}{7} 
\left(
\frac{\xi}{\pi} 
\right)^2 + \frac{15}{7} \left( \frac{\xi}{\pi} \right)^4
\right].
\end{equation}
So, if we define an effective number of massless neutrino families
$N_{eff} \equiv 3 + 30/7 (\xi / \pi)^2 + 15/7 ( \xi / \pi )^4$,
the mean density and pressure for all neutrinos will be given by these
ones for one massless non-degenerate family, multiplied by $N_{eff}$.

Let us now consider perturbed quantities.  We define $\Psi_{\nu}$ and
$\Psi_{\bar{\nu}}$, the perturbations of the phase space distribution for
$\nu$ and $\bar{\nu}$, through
\begin{eqnarray}
\delta f_{\nu} (\vec{x}, Q, \hat{n}, \tau) &=& f_{\nu} (Q) \Psi_{\nu} (\vec{x}, Q, \hat{n}, \tau)~, \
\nonumber \\
\delta f_{\bar{\nu}} (\vec{x}, Q, \hat{n}, \tau) &=& f_{\bar{\nu}} (Q) \Psi_{\bar{\nu}} (\vec{x}, Q, 
\hat{n}, \tau)
\end{eqnarray}
($\hat{n}$ is the momentum direction: $\vec{p} \equiv p \hat{n}$).
For our purpose, which is to integrate the linearized Einstein equations, 
it can be shown that only the following linear combination is relevant
\begin{equation}
\Psi \equiv \frac{ f_{\nu} \Psi_{\nu} + f_{\bar{\nu}} \Psi_{\bar{\nu}}}
{ f_{\nu} + f_{\bar{\nu}}}.
\end{equation}
Using the Boltzmann equations for $\Psi_{\nu}$ and $\Psi_{\bar{\nu}}$,
it is straightforward to show that the evolution of $\Psi$ (in Fourier
space and in the synchronous gauge, see \cite{MaBer}, eq. (40)) obeys
\begin{equation} \label{boltzmann}
\frac{\partial \Psi}{\partial \tau} +
i 
\frac{Q}{\sqrt{Q^2 + a^2 M^2}} 
(\vec{k}.\hat{n}) \Psi
+ 
\frac{d \ln ( f_{\nu} + f_{\bar{\nu}} )}{d \ln Q}
\left[ 
\dot{\eta} - 
\frac{\dot{h}+6\dot{\eta}}{2} 
(\hat{k}.\hat{n})^2 \right]=0~.
\end{equation}
This equation depends on $\xi$ only through the last term, which is
the gravitational source term. 

In the case $\xi=0$, the quantity $(d \ln ( f_{\nu} )/d \ln Q)$ has a
simple interpretation: it is the $Q$-dependence of a planckian
perturbation of the phase space distribution. In other words, a shift
of the blackbody temperature $\Delta T / T(\vec{x}, \hat{n}, \tau)$
corresponds to a perturbation
\begin{equation}
\Psi (\vec{x}, Q, \hat{n}, \tau) 
= - \frac{\Delta T}{T}(\vec{x}, \hat{n}, \tau) 
\frac{d \ln ( f_{\nu} )}{d \ln Q}. 
\end{equation}
Since the gravitational source term in the Boltzmann equation is
proportional to this quantity, the planckian shape is unaltered for
massless neutrinos, and also for massive neutrinos when they are still
relativistic (indeed, when $Q^2 \gg a^2 M^2$, the $Q$-dependence of
the Boltzmann equation (\ref{boltzmann}) vanishes in the second term,
and remains only in the third term).  When $\xi\neq0$, the source term
in eq. (\ref{boltzmann}) is proportional to
\begin{equation} \label{dlnfdlnq}
\frac
{d \ln ( f_{\nu} + f_{\bar{\nu}} )}
{d \ln Q} =
- 
\frac
{Q (1 + \ch \xi~\ch Q)}
{(\ch \xi + e^{-Q})(\ch \xi + \ch Q)}.
\end{equation} 
When neutrinos are still relativistic, $\Psi$ is proportional to
this quantity, even if it cannot be simply interpreted in
terms of blackbody temperature perturbations.

We can now specify all the changes required in {\tt cmbfast}, first in
the case of massive degenerate neutrinos. As usual, $\Psi$ can be
expanded in a Legendre series: $\Psi = \sum_{l=0}^{\infty} (-i)^l (2l
+1) \Psi_l P_l$.  It is easy to show that for each multipole $\Psi_l$,
the evolution equation and the initial condition are both identical to
those of the non-degenerate case, provided that we replace $(d \ln (
f_{\nu} )/d \ln Q)$ by eq. (\ref{dlnfdlnq}).  So, in summary, one only
needs to modify the homogeneous phase-space distribution, its
logarithmic derivative with respect to $Q$, and the initial
calculation of $\Omega_{\nu}$.  Also, in order to obtain a good
precision in the CMB anisotropy spectra, one must set $l=5$ for the
number of multipoles $\Psi_l$ to be time-integrated.  For transfer
functions, the value $l=25$ proposed by the code is sufficient.

In the case of massless degenerate neutrinos, the situation is even
simpler.  The $Q$-dependence of the Boltzmann equation can be
integrated away, just like in the non-degenerate case. For this
purpose, we must introduce the $Q$-independent variable $F_{\nu}$
\begin{equation}
F_{\nu} (\vec{k}, \hat{n}, \tau) \equiv
\frac{\int Q^3 dQ (f_{\nu} + f_{\bar{\nu}}) \Psi} 
{\int Q^3 dQ (f_{\nu} + f_{\bar{\nu}})}
\equiv \sum_{l=0}^{\infty} (-i)^l (2l+1) F_{\nu l} P_l,
\end{equation}
and integrate by part the last term in equation (\ref{boltzmann}).
The multipoles $F_{\nu l}$ are exactly identical for degenerate and
non-degenerate massless neutrinos, because they share the same
evolution equations and initial conditions. So, the effect of $\xi$
arises only through the background quantities in \eq{background}
and is completely described by introducing an effective number of massless
neutrinos. 

\section{Results}
\label{results}

First, as a consistency check, we compute CMB anisotropies and
transfer functions for different values of $\xi$, choosing a very
small mass $m_{\nu} \leq 0.001$ eV. We check that the results match
exactly those obtained with the unmodified version of {\tt cmbfast},
when the appropriate effective neutrino number $N_{eff}$ is specified.

The effect of $\xi$ and $m_{\nu}$ on the CMB anisotropy spectrum can
be seen in figure \ref{fig.CMB}. We choose a set of cosmological
parameters ($h=0.65$, $\Omega_b=0.05$, $\Omega_{\Lambda}=0.70$,
$\Omega_{CDM}=1-\Omega_b-\Omega_{\nu}-\Omega_{\Lambda}$,
$Q_{rms-ps}=18~\mu$K, flat primordial spectrum, no reionization, no
tensor contribution), and we vary $\xi$ from 0 to 5, both in the case
of massless degenerate neutrinos (solid lines) and degenerate
neutrinos with $m_{\nu}=0.07$ eV (dashed lines).  Let us first comment
the massless case.  The main effect of $\xi$ is to boost the amplitude
of the first peak\footnote{In fact, this is not true for very large
values of $\xi$.  In such cases, recombination can take place still at
the end of radiation domination, and anisotropies are suppressed. For
our choice of cosmological parameters, this happens for $\xi \gsim 7$,
but in such a case the location of the first peak is $l \gsim 450$, and
the matter power spectrum is strongly suppressed.}.  Indeed,
increasing the energy density of radiation delays matter-radiation
equality, which is known to boost the acoustic peaks \cite{Hu} (the
same explanation holds for the effect of $\Omega_{\Lambda}$ in flat
models). For the same reason, all peaks are shifted to higher
multipoles, by a factor $( (1 + a_{eq}/a_*)^{1/2} -
(a_{eq}/a_*)^{1/2})^{-1}$ \cite{Hu} ($a_{eq}$ is the scale factor at
equality, and increases with $\xi$, while the recombination scale
factor $a_*$ is almost independent of the radiation energy density).
Secondary peaks are then more affected by diffusion damping at large
$l$, and their amplitude can decrease with $\xi$.

In the case of degenerate neutrinos with $m_{\nu}=0.07$ eV, the
results are quite similar in first approximation. Indeed, the effects
described previously depend on the energy density of neutrinos at
equality. At that time, they are still relativistic, and identical to
massless neutrinos with equal degeneracy parameter.  However, with a
large degeneracy, $\Omega_{\nu}$ today becomes significant: for
$\xi=5$, one has $\Omega_{\nu}=0.028$, i.e. the same order of
magnitude as $\Omega_b$.  Since we are studying flat models,
$\Omega_{\nu}=0.028$ must be compensated by less baryons, cold dark
matter (CDM) or $\Omega_{\Lambda}$. In our example, $\Omega_b$ and
$\Omega_{\Lambda}$ are fixed, while $\Omega_{CDM}$ slightly
decreases. This explains the small enhancement of the first peak
compared to the massless case.  Even if this effect is indirect, it is
nevertheless detectable in principle (even if one does not impose the
flatness condition, the effect of $\Omega_{\nu}$ will be visible
through a modification of the curvature). In figure \ref{fig.CMB}, for
$\xi=0$, the first peak maximum is enhanced by only 0.37\%, while for
$\xi=5$, there is an increase of 3.4\%, detectable by the future
satellite missions {\it MAP} and {\it Planck}, unless there are large
parameter degeneracies.  It is well-known that such degeneracies are
generally removed when CMB and LSS data are combined for parameter
extraction \cite{Tegmark}.

We plot in figure \ref{fig.PK} the power spectrum $P(k)$ for the same
models as in figure \ref{fig.CMB}, normalized on large scales to
COBE. The effect of both parameters $\xi$ and $m_{\nu}$ is now to
suppress the power on small scales.  Indeed, increasing $\xi$
postpones matter-radiation equality, allowing less growth for
fluctuations crossing the Hubble radius during radiation
domination. Adding a small mass affects the recent evolution of
fluctuations, and has now a direct effect: when the degenerate
neutrinos become non-relativistic, their free-streaming suppresses
growth of fluctuations for scales within the Hubble radius.  For
non-degenerate neutrinos, this effect is known to reduce power on
those scales by a relative amount $\Delta P/P \sim 8 \Omega_{\nu} /
\Omega_0$ \cite{Huetal} (we introduced $\Omega_0 = 1 -
\Omega_{\Lambda}$). So, even with $m_{\nu}=0.07$ eV and $\xi=0$, it is
significant, especially at low $\Omega_0$. In the models of figure
\ref{fig.PK}, $P(k)$ decreases by $\sim 5$ \%, in agreement with the
theoretical prediction ($\Omega_{\nu}=1.8 \times 10^{-3}$,
$\Omega_0=0.3$). However, at $\xi=5$ (i.e.  $\Omega_{\nu}=0.028$),
this effect is even larger: $P(k)$ decreases by a factor 2.2, instead
of an expected 1.7. This effect is likely related to the phase-space
distribution of neutrinos with a chemical potential: their average
momentum is shifted to larger values, making the free-streaming
suppression mechanism even more efficient.

Let us compare our results with those of previous works. The effect of
$\xi$ on the CMB for massless neutrinos and $\Omega_{\Lambda}=0$ is
the same as that one found in \cite{Sarkar}.  We also agree with the
revised results in \cite{Kinney}.

\section{Comparison with observations}
\label{comparison}

Since the degeneracy increases dramatically the amplitude of the first
CMB peak, we expect large $\xi$ values to be disfavored in the case of
cosmological models known to predict a fairly high peak.  On the other
hand, a high $\xi$ is likely to be allowed (or even favored) for
models that predict systematically a low peak, unless a large scalar
spectral index $n \geq 1.2$ ({\it blue tilt}) is invoked. For
instance, the degeneracy is likely to be favored by: (i) a large
contribution of tensor perturbations; (ii) a significant effect from
reionization; (iii) a low baryon density; (iv) a large $h$ ($h \geq
0.7$); (v) flat models with $\Omega_{\Lambda} \leq 0.6$; etc.  For
such models, the peak amplitude can be boosted by $\xi$, keeping $n$
close to one, which is more natural from the point of view of
inflation. However, a careful case-by-case analysis is required, since
the effects of $\xi$ and $n$ on CMB and LSS spectra are far from being
equivalent.  Our goal here is not to explore systematically all
possibilities, but to briefly illustrate how $\xi$ can be constrained
by current observations for flat models with different values of
$\Omega_{\Lambda}$. Recent results from supernovae
\cite{Perlmutter}, combined with CMB constraints, favor flat models
with $\Omega_{\Lambda} \sim 0.6-0.7$ \cite{Lineweaver}.

We choose a flat model with $h=0.65$, $\Omega_b=0.05$,
$Q_{rms-ps}=18~\mu$K, no reionization and no tensor contribution, and
look for the allowed region in the space of free parameters
($\Omega_{\Lambda},\xi,n$).  The allowed region will not be defined
using a maximum likelihood analysis, but with the more conservative
technique called ``concordance'' by Wang et al. \cite{Wang}, which
consists in taking the intersection of regions allowed by each
experiment.

For simplicity, we take into account only a few constraints on the
matter power spectrum, known to be representative of the large amount
of available data: the value of $\sigma_8$ (the variance of mass
fluctuations in a sphere of radius $R=8 h^{-1} $Mpc) given for flat
models in \cite{ViaLid}, at 95 \% confidence level (CL)\footnote{The
Viana \& Liddle result \cite{ViaLid} is in very good agreement with an
independent derivation by Girardi et al. \cite{Borgani}.}; a
$\chi^2$ comparison with the STROMLO-APM redshift survey
\cite{Loveday}, at scales well within the linear regime, also with 95
\% CL\footnote{This confidence level stands for the goodness-of-fit of
the model: when $\chi^2$ is greater than some value, the probability
that we find the observed dataset, assuming the model to be valid, is
smaller than 5 \%. For the APM data, we have $9-3$ degrees of freedom,
and the limiting value is found in numerical tables to be
$\chi^2=12.5$.}; and finally, the constraint on bulk velocity at $R=
50 h^{-1}$ Mpc \cite{KolDek}, taking into account the cosmic variance.
Except for the updated $\sigma_8$ constraint, we use exactly the same
experimental tests as in \cite{LPS}, and refer the reader to this
paper for details. For CMB data, we perform a $\chi^2$ analysis based
on 19 experimental points and window functions, taking into account
the Saskatoon calibration uncertainty, in the way suggested by
\cite{LineBar}. The list of data that we use is given in \cite{LPS},
and again allowed regions correspond to 95 \% CL\footnote{Here we have
$19-3$ d.o.f.; the 95 \% CL is given by $\chi^2 = 26$.}.  We do not
take into account the most recent experiments, for which window
functions are still unpublished; they are anyway in good agreement
with the data considered here.

We plot in figure \ref{fig.WIN} the LSS and CMB allowed regions in
($\xi$, $n$) parameter space, corresponding to
$\Omega_{\Lambda}=0,0.5,0.6,0.7$.  For $\Omega_{\Lambda}=0.5-0.7$, the
LSS window just comes out of $\sigma_8$ limits.  For
$\Omega_{\Lambda}=0$, the lower LSS constraint is from $\sigma_8$, and
the upper one from APM data. In the case of degenerate neutrinos with
$m_{\nu} = 0.07$ eV, the windows are slightly shifted at large $\xi$,
since, as we saw, the effect of $\xi$ is enhanced (dotted lines on the
figure). The CMB allowed regions do not show this distinction,
given the smallness of the effect and the imprecision of the data.
One can immediately see that LSS and CMB constraints on $n$ are shifted in
opposite direction with $\xi$: indeed, the effects of $\xi$ and $n$
both produce a higher CMB peak, while to a certain extent they
compensate each other in $P(k)$.  So, for $\Omega_{\Lambda}=0.7$, a
case in which a power spectrum normalized to both COBE and $\sigma_8$
yields a too high peak\footnote{At least, for the values of the other
cosmological parameters considered here.  This situation can be easily
improved, for instance, with $h=0.7$.}, a neutrino degeneracy can only
make things worst.  On the other hand, for $\Omega_{\Lambda}=0.5-0.6$,
a good agreement is found up to $\xi \simeq 3$.

Let us finally consider the $\Omega_{\Lambda}=0$ case in which, after
COBE normalization of the power spectrum, there is a well known
discrepancy between the amplitude required by $\sigma_8$ and the shape
probed by redshift surveys: these two constraints favor different
values of $n$. We find that the neutrino degeneracy can solve this
problem with $\xi \gsim 3.5$; however, the allowed window is cut at
$\xi \simeq 6$ by CMB data, and we are left with an interesting region
in which $\Omega_0=1$ models are viable. This result is consistent
with \cite{Sarkar}. However, current evidences for a low $\Omega_0$
Universe \cite{Perlmutter,Bahcall} are independent of the constraints
used here, so there are not many motivations at the moment to consider
this window seriously.

\section{Conclusions}

We have considered some cosmological implications of a large relic
neutrino degeneracy. We have shown that this degeneracy enhances the
contribution of massive neutrinos to the present energy density of the
Universe. For instance, neutrinos with a small mass $m_{\nu} \sim
10^{-2}$ eV can contribute significantly to $\Omega_0$, provided that
there is a large neutrino-antineutrino asymmetry.

Our main result is the computation of the power spectra of CMB
anisotropies and matter density in presence of a neutrino
degeneracy. We found, in agreement with \cite{Sarkar}, that it boosts
the amplitude of the first CMB peak, shifts the peaks to larger
multipoles, and supresses small scale matter fluctuations.  These
effects follow the increase of neutrino energy density, that delays
matter-radiation equality.

We extended the calculation to the case of massive degenerate
neutrinos, and showed the results for a mass of $0.07$ eV, as
suggested by the Super-Kamiokande experiment. This mass has a small
effect on CMB anisotropies. Indeed, such light neutrinos are still
relativistic at recombination, but in presence of a degeneracy, they
can account for a substantial part of the density today, of order
$\Omega_{\nu} \sim 10^{-2}$.  Also, we showed that small scale matter
fluctuations are much more suppressed when the degenerate neutrinos
are massive, because free-streaming of non-relativistic neutrinos is
more efficient when their average momentum is boosted by the chemical
potential.

We compared our results with observations, in the restricted case of a
flat universe with arbitrary ($\Omega_{\Lambda}, \xi, n$) and fixed
values of other cosmological parameters. We found that for
$\Omega_{\Lambda} \simeq 0.5 - 0.6$, a large degeneracy is allowed, up
to $\xi \simeq 3$. However, this upper bound is smaller than the value
$\xi \simeq 4.6$ needed to explain the generation of ultra-high energy
cosmic rays by the annihilation of high-energetic neutrinos on relic
neutrinos with mass $m_{\nu}= 0.07$ eV \cite{Gelmini}.  We also tried
smaller values of $\Omega_{\Lambda}$, even if they are not favored by
combined CMB and supernovae data. It turns out that a large degeneracy
can account for both CMB and LSS constraints even with $\Omega_0 =1$,
provided that $3.5 \leq \xi \leq 6$.  This analysis could be extended
to other cosmological models. For instance, the degeneracy is likely
to be compatible with a large contribution of tensor perturbations to
large scale CMB anisotropies.

Finally, it turns out that the degeneracy parameter and the mass of
degenerate neutrinos have effects within the level of detectability of
future CMB observations and redshift surveys, even with $m_{\nu} \sim
0.07$ eV.  However, a careful analysis should be performed in order to
detect possible parameter degeneracy between $\xi$, $m_{\nu}$ and
other cosmological parameters.

\section*{Acknowledgements}

We thank S.~Borgani, A.~Masiero and A.Yu.~Smirnov for useful
discussions, as well as W.~Kinney and A.~Riotto for correspondence.
This work has been supported by INFN and by the TMR network grant
ERBFMRXCT960090.

\newpage
\begin{figure}
\caption{Present energy density of massive degenerate neutrinos as a
function of the degeneracy $\xi$. The curves correspond to
different values of $h^2 \Omega_\nu$ and the horizontal line is the upper 
bound from \eq{lsf}.}
\label{ximass}
\end{figure}
\begin{figure}
\caption{Same as figure \ref{ximass} for the neutrino asymmetry $L_\nu$.}
\label{lnumass}
\end{figure}
\begin{figure}
\caption{CMB anisotropy spectrum for different models with one family
of degenerate neutrinos. Solid lines account for the case of massless
degenerate neutrinos, and correspond, from bottom to top, to $\xi=0,3,5$.
Dashed lines correspond to degenerate neutrinos with mass $m_{\nu} = 0.07$ eV.
Other parameters are fixed to $h=0.65$, $\Omega_b=0.05$, $\Omega_{\Lambda}=0.70$,
$\Omega_{CDM}=1-\Omega_b-\Omega_{\nu}-\Omega_{\Lambda}$, $Q_{rms-ps}=18~\mu$K,
$n=1$. We neglect reionization and tensor contribution.}
\label{fig.CMB}
\end{figure}
\begin{figure}
\caption{Present power spectrum of matter density, 
for the same parameters as in the previous figure. {}From
top to bottom, to $\xi=0,3,5$.}
\label{fig.PK}
\end{figure}
\begin{figure}
\caption{LSS and CMB constraints in ($\xi$, $n$) space, for different
choices of $\Omega_{\Lambda}$: from top left to bottom right,
$\Omega_{\Lambda}=0,0.5,0.6,0.7$.  The underlying cosmological model
is flat, with $h=0.65$, $\Omega_b=0.05$, $Q_{rms-ps}=18~\mu$K, no
reionization, no tensor contribution. The allowed regions are
those where the labels are. For LSS constraints, we can
distinguish between degenerate neutrinos with $m_{\nu} =0$ (solid
lines) and $m_{\nu} =0.07$ eV (dotted lines).}
\label{fig.WIN}
\end{figure}
\newpage
\thispagestyle{empty}
\centerline{\psfig{file=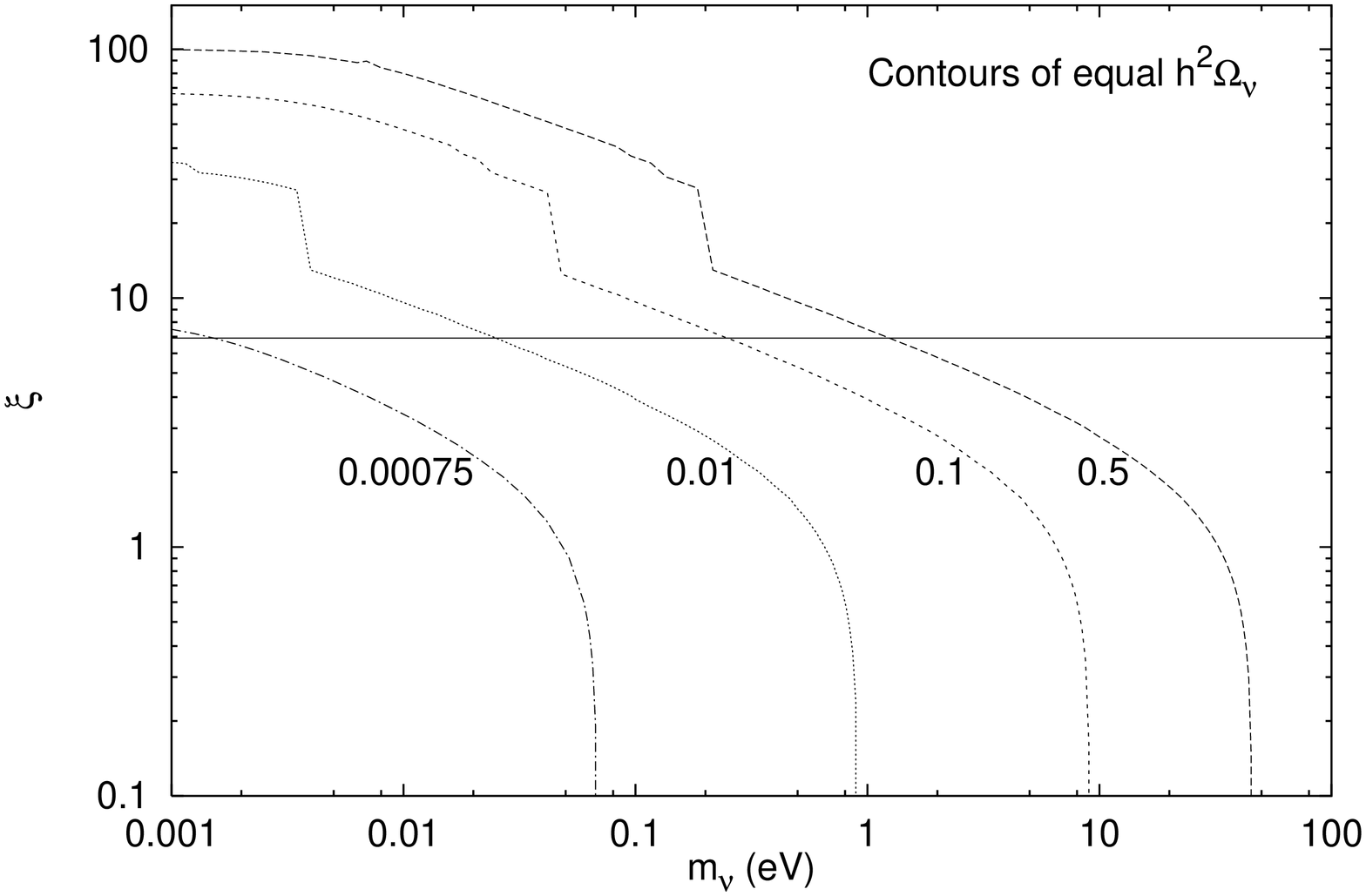,angle=-90,width=0.9\textwidth}}
\vspace{2cm}
\centerline{Figure 1}
\newpage
\thispagestyle{empty}
\centerline{\psfig{file=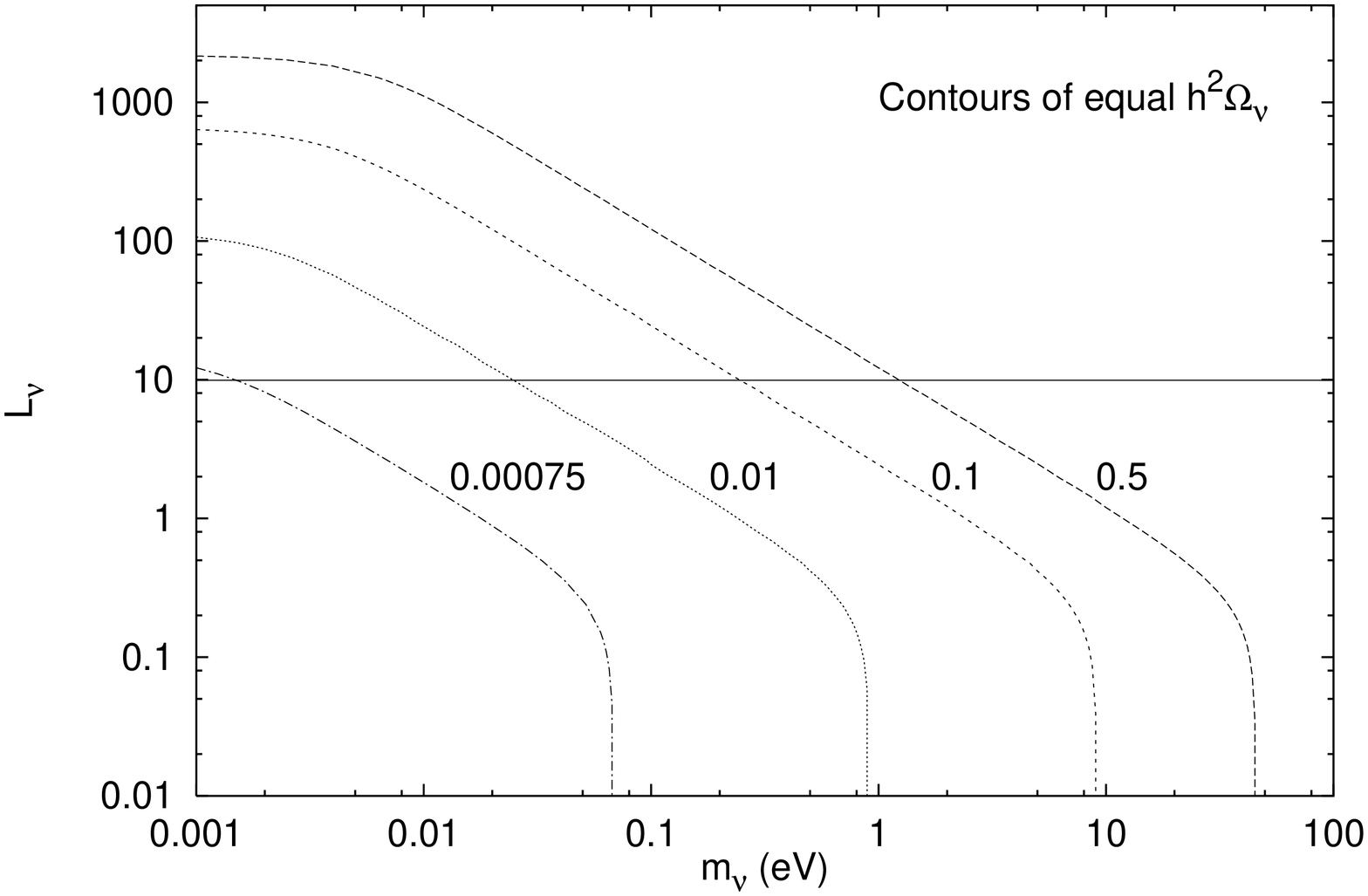,angle=-90,width=0.9\textwidth}}
\vspace{2cm}
\centerline{Figure 2}
\newpage
\thispagestyle{empty}
\centerline{\psfig{file=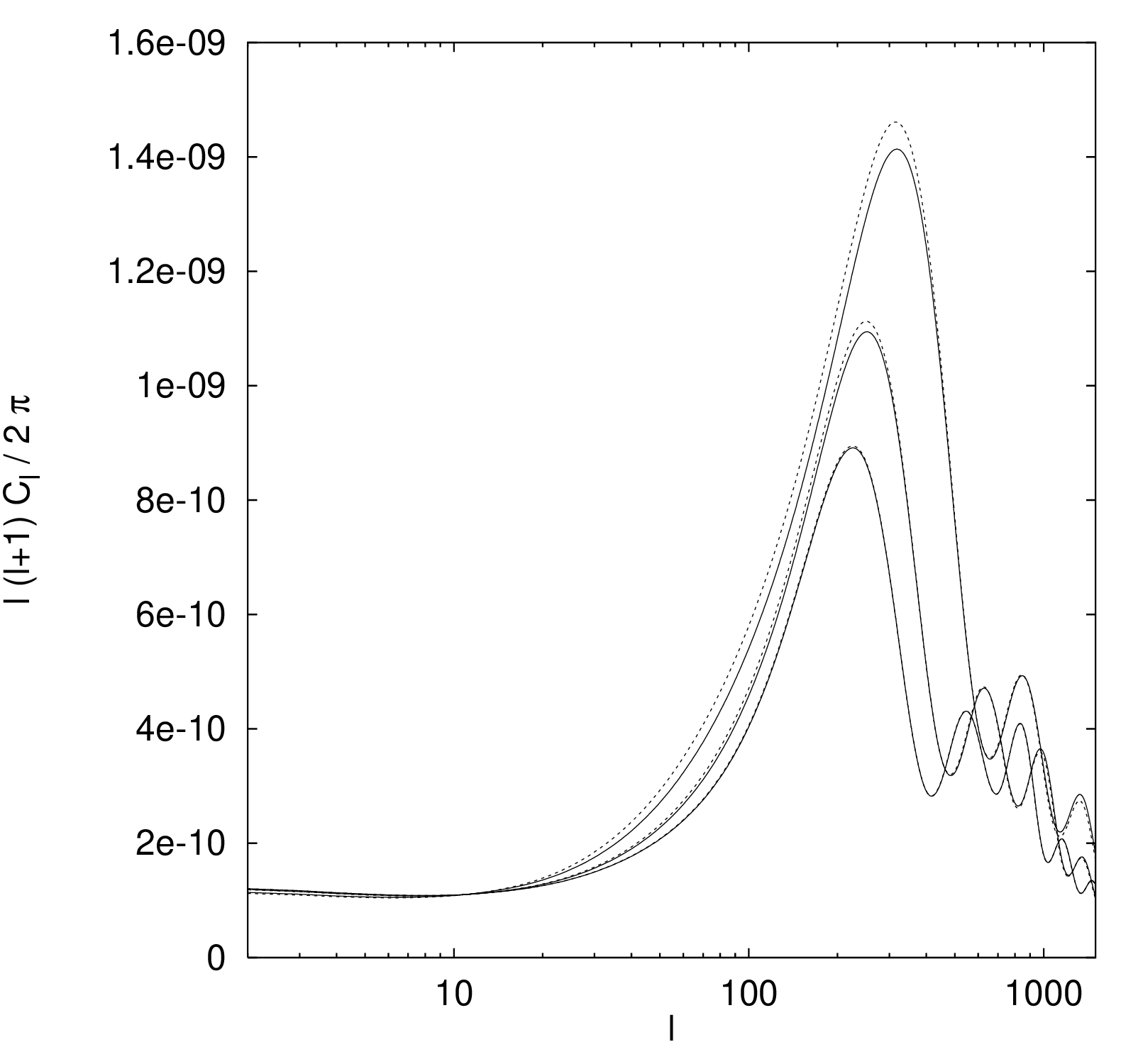,width=0.8\textwidth}}
\vspace{2cm}
\centerline{Figure 3}
\newpage
\thispagestyle{empty}
\centerline{\psfig{file=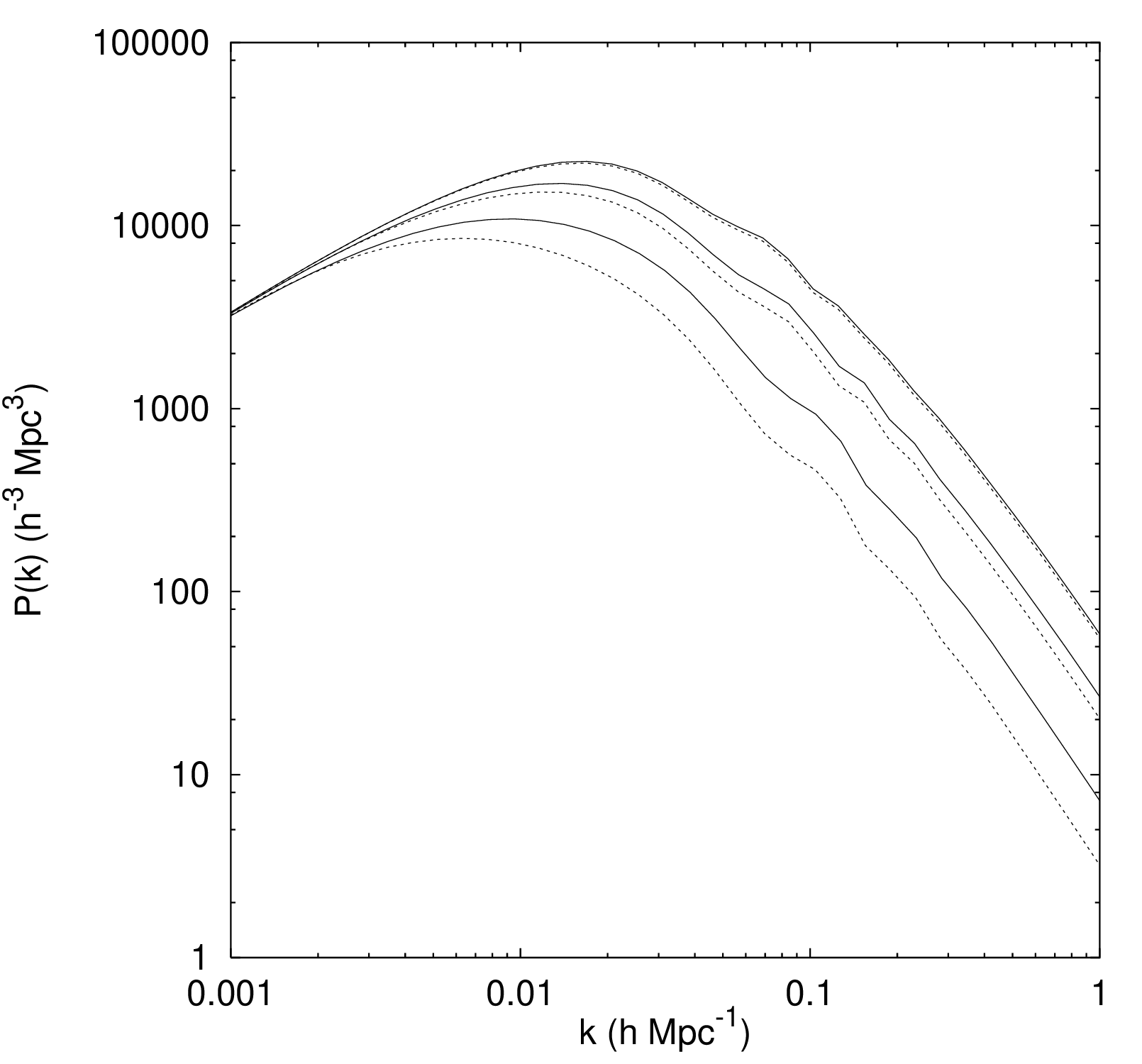,width=0.8\textwidth}}
\vspace{2cm}
\centerline{Figure 4}
\newpage
\thispagestyle{empty}
\begin{eqnarray}
\epsfysize=8cm
\epsfbox{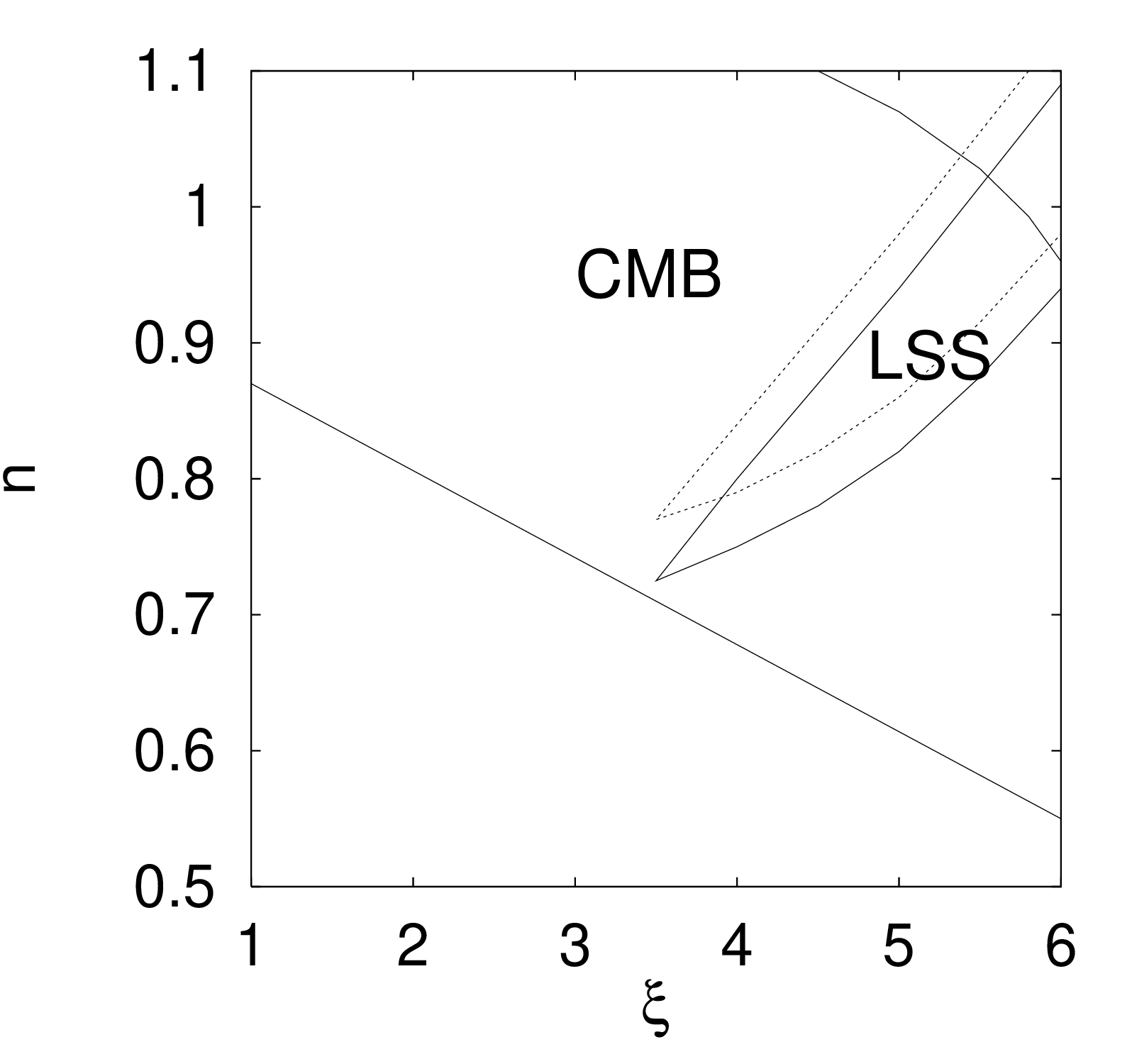}
&&
\epsfysize=8cm
\epsfbox{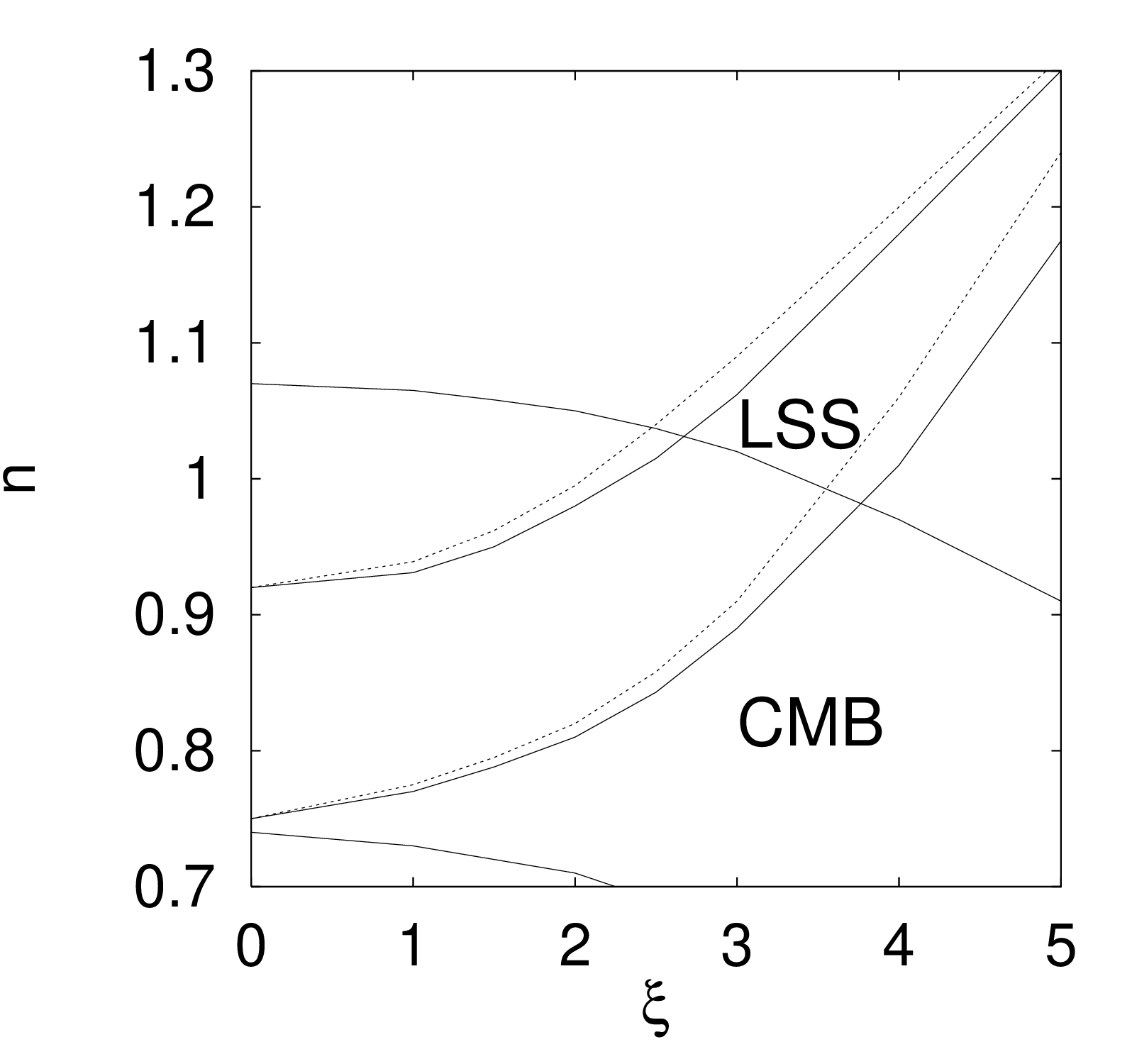}
\nonumber \\
\epsfysize=8cm
\epsfbox{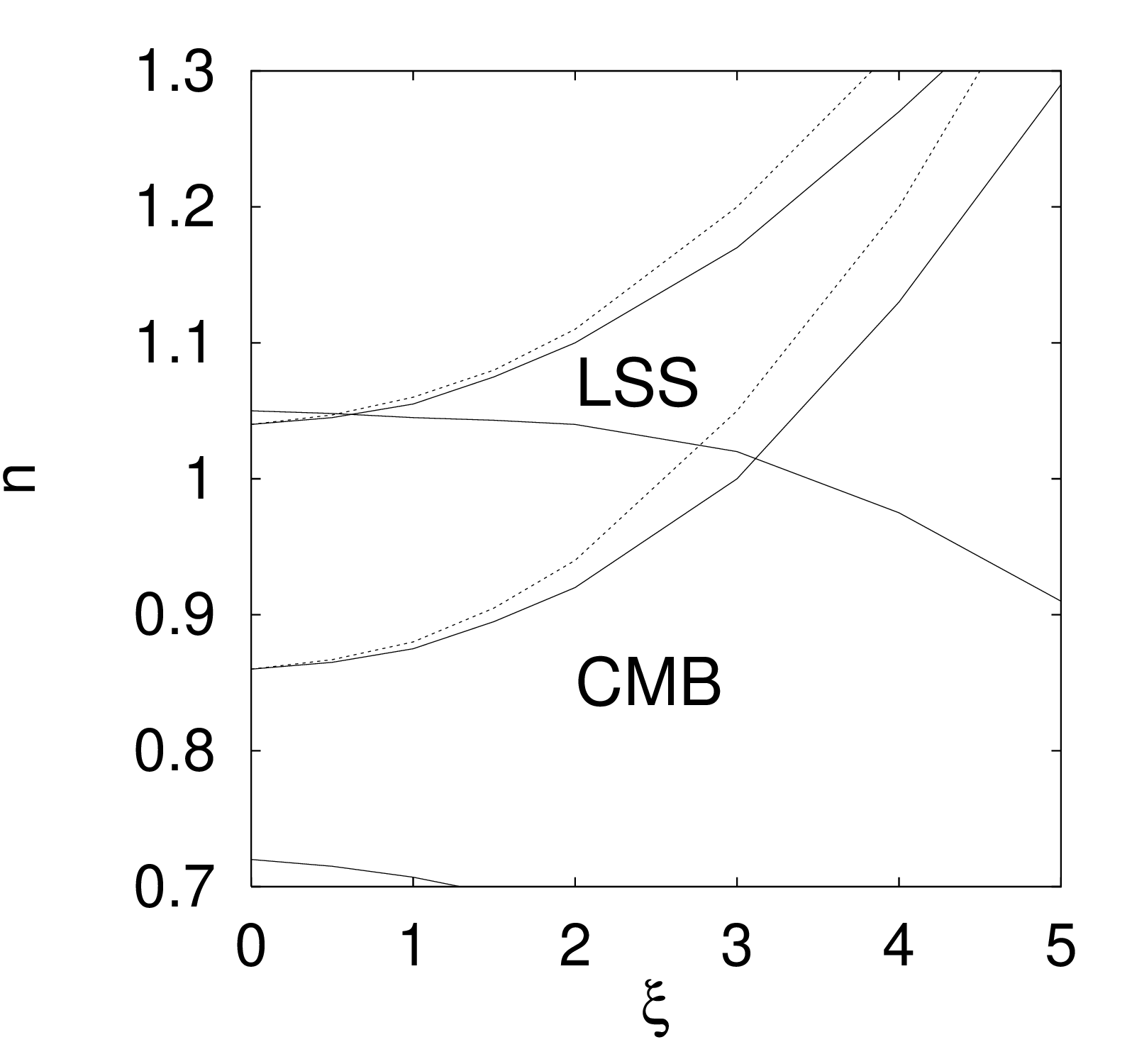}
&&
\epsfysize=8cm
\epsfbox{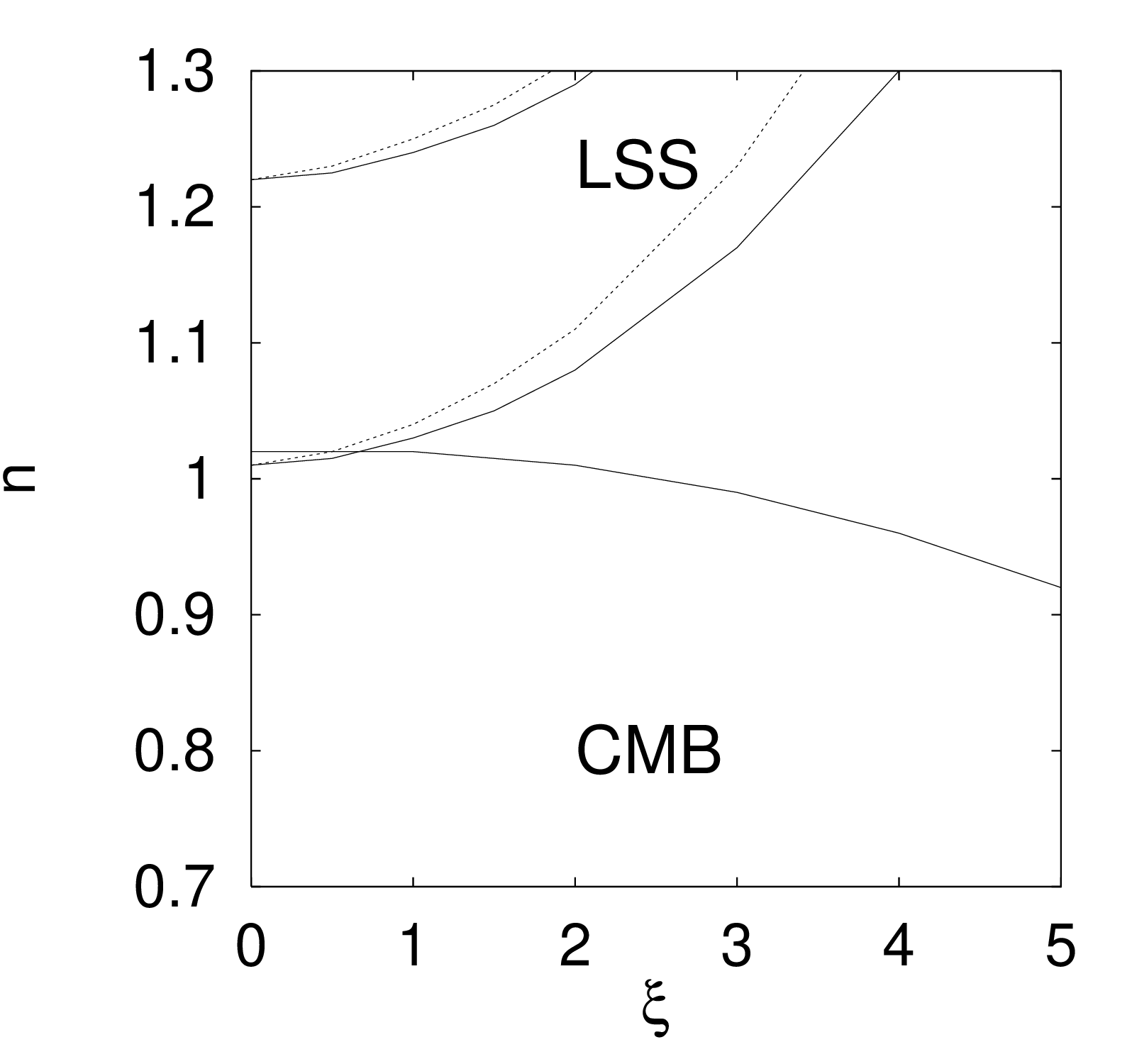}
\nonumber
\end{eqnarray}
\vspace{1cm}
\centerline{Figure 5}
%

\end{document}